# Arbitrarily polarized and unidirectional emission from thermal metasurfaces

J. Ryan Nolen[1]*, Adam C. Overvig[1]*, Michele Cotrufo[1], and Andrea Alù[1,2]†

[1]Photonics Initiative, Advanced Science Research Center, City University of New York, New York, NY 10031, USA

[2]Physics Program, Graduate Center of the City University of New York, New York, NY 10016, USA

*These authors contributed equally. †Corresponding author: aalu@gc.cuny.edu

**Abstract:**

*Thermal emission from a hot body is ubiquitous, yet its properties remain inherently challenging to control due to its incoherent nature. Recent advances in thermal emission manipulation have been unveiling exciting phenomena and new opportunities for applications. In particular, judiciously patterned nanoscale features over their surface have been shown to channel emission sources into partially coherent beams with tailored directionality and frequency selectivity. Yet, more sophisticated forms of control, such as spin-selective and unidirectional thermal emission have remained elusive. Here, we experimentally demonstrate single-layer metasurfaces emitting unidirectional, narrowband thermal light in the infrared with arbitrary polarization states – an operation enabled by photonic bound states in the continuum locally tailored by a geometric phase controlling the temporal and spatial coherence of emitted light. The demonstrated platform paves the way to a compactification paradigm for metasurface optics, in which thermal emission or photoluminescence can feed arbitrarily patterned beams without the need of external coherent sources.*

Incandescence is a ubiquitous source of light, yielding highly incoherent, broadband, uncollimated and unpolarized emission [1]. These features often limit its applicability and broad appeal compared to coherent sources such as lasers. Addressing these shortcomings may pave the way to a broad range of cheap infrared sources with groundbreaking impact on various technologies [2]. Towards this goal, recent demonstrations have been leveraging strongly coupled material resonances with surface waves to enhance the temporal and spatial coherence of thermal emission, leading to narrowband, collimated and linearly polarized responses [3]-[12]. These structures typically rely on phonon polaritons [13], which naturally support strong coupling of mid-infrared light with lattice vibrations, demonstrating narrow linewidth peaks in thermal emission [14]. However, these features are limited to specific material platforms and frequencies, being spectrally bound by the optical phonons of the available materials. Such materials are not necessarily compatible with integrated technologies (which heavily favor silicon and its oxide), and in particular have left a large part of the mid-infrared (IR) range (including the $3-5\ \mu m$ transparency window) inaccessible. More generally, this surface-wave-mediated emission has thus far been limited to simple, symmetric forms of wavefront shaping to *p*-polarized waves; generation of beams at arbitrary optical frequencies with arbitrary spatial profiles of amplitude, phase and polarization still requires complex and bulky optical systems comprising cascaded stacks of optical devices, which comes at the cost of footprint and efficiency. In parallel, engineered thin films—often termed 'metasurfaces'—have been introducing a paradigm of compactification of optical systems, realizing a wide range of functionalities including holography, beam steering and focusing over an ultrathin platform. However, these devices are designed under the assumption of being driven by externally produced coherent radiation, typically from bulky and expensive optical sources [15]-[16].

The fields of thermal emission and optical metasurfaces have been converging in recent times with the discovery of new emergent phenomena stemming from careful engineering of the optical density of states. For instance, the photonic Rashba effect can support bi-directional thermal emission of circular polarization (CP) beams with spin splitting, expanding the capabilities of structured surfaces to act as both the element and the source. However, these devices only function at oblique angles, and do not offer independent control over the CP states or other forms of polarization selectivity [17]-[18]. Symmetry-broken metasurfaces have recently demonstrated control over emission and absorption to include circular polarization states at normal incidence [19],[20], though these demonstrations have been limited to lossy metallic structures, restricting the ultimate spatial and temporal coherence as well as manufacturing compatibility. Relatedly, the incorporation of metasurface concepts into luminescent structures has enabled directional emission through one of the two half-spaces [21],[22], but so far is limited in polarization and directivity. Despite these recent successes and extensive efforts to tackle the control over thermal emission and photoluminescence, fundamental limits remain in the command over light stemming from incoherent oscillations of matter. For instance, asymmetric, unidirectional thermal emission of a chosen arbitrary polarization to a single half-space remains elusive. Recently, thermal metasurfaces [23], based on the emerging principles of diffractive nonlocal engineering in nano-optics [24]-[25], have been theoretically put forward, demonstrating that they can support spin-selective, unidirectional emission, focused emission, and wavefront patterning, e.g., producing angular orbital momentum thermal beams. In this initial theoretical proposal, the implementation requires multi-layer nanofabrication, demanding careful alignment, and hence presenting a major barrier to their experimental realization. A single-layer monolithic implementation of thermal metasurfaces with high coherence using standard, low-loss dielectric materials compatible with CMOS platforms remains a key unsolved obstacle to enabling the promise of thermal metasurfaces, i.e., the generation of custom, highly coherent wavefronts at a wavelength of choice by simply heating an ultrathin nanostructured film.

Here, we implement symmetry-controlled nonlocal thermal metasurfaces with customizable chiral emission mediated by a local geometric phase (Fig. 1(a)). We realize our metasurface with amorphous silicon (a-Si) pillars atop silicon dioxide and gold films, with four elements per unit cell designed to control the absorption/emission, mediated by nonlocal scattering from the interfaces of the $a$-Si layer (Fig. 1(b)). The design is based on the implementation of a quasi-bound state in the continuum (q-BIC) [26] supported by a periodic array of pillars, whose in-plane long-range order provides coherence to the thermal emission (Fig. 1(c)) [3], producing narrowband, directional CP thermal emission. By varying the two orientation angles $\theta$ and $\alpha$ with respect to the fixed lattice, nearly full coverage over the entire Poincaré sphere can be enabled (Fig. 1(d)) at normal incidence. Intrinsic to this polarization control is a local orientability of the linear polarizations comprising the CP states. In turn, this local control introduces a geometric phase (Fig. 1(e), inset), a manifestation of the fundamental symmetry of an infinite helix, i.e., a rotation about the axis is equivalent to translation along the axis (akin to an Archimedes' screw). We exploit this phase to construct a library of meta-units maintaining a high degree of CP (DoCP) across the entire device, while simultaneously imprinting an arbitrary local pattern of geometric phase

$\Phi_{geo}$ spanning $2\pi$ (Fig. 1(e)). By symmetry, the values $\alpha = (90°, 180°]$ are equivalent to the set $\alpha = (0°, 90°]$ but with a $\pi$ phase shift arising from a sign flip in the q-BIC scattering [24].

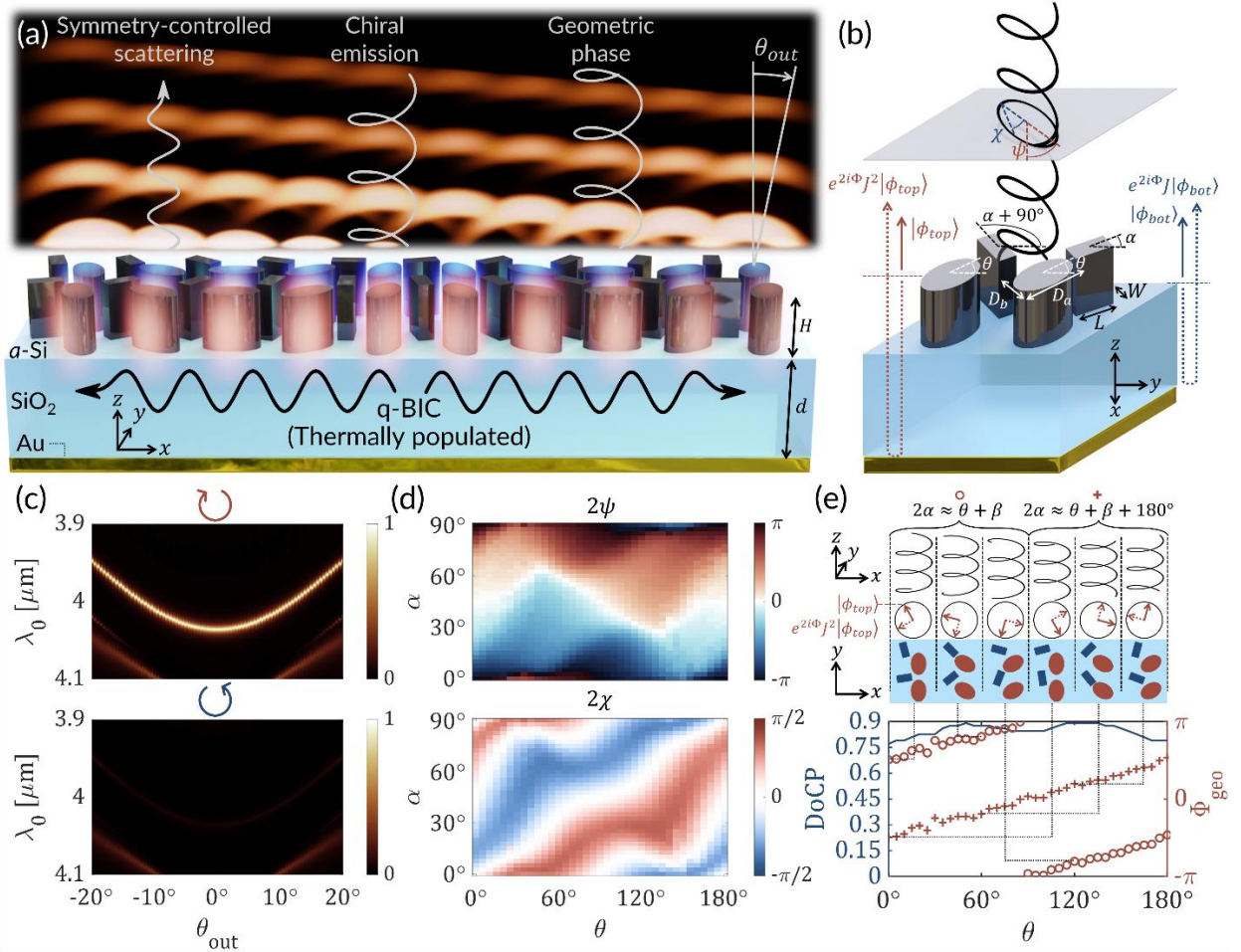

**Fig. 1. Design and functionality of a thermal metasurface.** (a) Schematic view of the symmetry-controlled scattering mediated by a thermally populated q-BIC, exhibiting chiral emission with a tunable geometric phase. (b) Geometry of a meta-unit cell, with depictions of the scattered components from the top and bottom interfaces of the silicon layer. (c) Simulated RCP (top) and LCP (bottom) emission from an example meta-unit array. (d) Simulated polarization emitted by meta-units varying $\theta$ and $\alpha$. (e) Simulated DoCP and geometric phase $\Phi_{geo}$ (bottom) of a selected set of meta-units, with example characteristic polarization helixes (top).

Based on these principles, we fabricated six different metasurface devices to demonstrate full control over the polarization state across the entire Poincaré sphere within a single metasurface platform (Fig. 2). We stress that this is the first report of a single platform capable of exhibiting thermal emission with arbitrarily specified elliptical polarization, and that the Q-factors observed here (ranging from $Q = 100$ to $Q = 200$) are substantially higher than those of the plasmonic approaches this spectral range is conventionally limited to (Q~20-25). Next, we fabricated a metasurface encoding a geometric phase gradient exhibiting spin-selective unidirectional emission to a single half-space (Fig. 3). In direct contrast to commonly observed spin-splitting phenomena, wherein CP of one handedness emitting to one direction necessitates the presence of CP of the opposite handedness emitting to the opposite direction, this device experimentally demonstrates for the first time that reciprocal systems may control a single spin at will, without the presence of the opposite spin. Quite remarkably, we achieve this not only without the need of breaking reciprocity, but also within a single layer metasurface platform over a substrate. We confirm these concepts through a novel spatio-temporal coupled mode theory (STCMT) model, numerical simulations, and experimental characterization of our devices.

The metasurfaces are designed to operate around $4\mu m$, within the $3 - 5\ \mu m$ atmospheric transparency window typically inaccessible by thermal structures mediated by phonon polariton resonances. The emission can be tuned throughout the mid-IR through a simple scaling of the geometric parameters (see Supplementary Information), since we use nondispersive, CMOS compatible dielectric materials to implement these phenomena. The meta-units comprise four micropillars: two identical *local* pillars (elliptical) oriented at an angle $\theta$, controlling the local phase, and two *nonlocal* pillars (rectangular) oriented at angles $\alpha$ and $\alpha + 90°$, supporting the extended q-BIC mode. The angles $\theta$ and $\alpha$ are defined as the counterclockwise in-plane angles between the elongated direction of the corresponding pillar and the $y$-axis. The rectangular bars lie at the interstitial sites of the monomer elliptical array, as shown in Fig. 1(b). The local pillars' major and minor diameters, $D_a$ and $D_b$, are tuned to ensure a birefringent response, thus conferring a linear control over the geometric phase tailoring the parameter $\theta$ [16]. In contrast to conventional geometric phase metasurfaces, these pillars at the same time support a transverse magnetic q-BIC whose modal profile promotes invariance in Q-factor and resonant frequency across the entire device, despite the local parameter $\theta$ being varied. When the length $L$ and width $W$ of the nonlocal pillars are equal, this state is symmetry protected, being bound to the surface and hence emitting no thermal radiation. As the symmetry in the nonlocal pillars is broken, by making $L$ and $W$ unequal, the bound state becomes a q-BIC, and leaks to free-space with a radiative Q-factor $Q \propto (L-W)^{-2}$ [26]. These parameters are tuned such that the radiative and nonradiative Q-factors are identical across the entire device, a critical coupling condition that enables black-body thermal emissivity with a linewidth customizable by tuning both the amount of material loss and the perturbation strength [27].

We control the scattering by the nonlocal pillars following selection rules that govern the excitation of q-BICs, which in vertically-invariant (extruded) structures are limited to linear polarizations generated at both the top and bottom interfaces of the silicon pillars [24]. While previous theoretical work achieved chiral q-BICs by manipulating the two interfaces independently [25] or by introducing the nonlocal pillars only at the bottom interface [23], here we aim for a single layer geometry to facilitate the fabrication and experimental demonstration. To address this challenge, we engineer the silicon thin film's average refractive index and the silicon dioxide thin film's thickness in order to largely suppress the scattering off the bottom interface via destructive interference. Consequently, the scattering from the top interface dominates the selection rules, and the emitted polarization state is the superposition of (i) the *directly* scattered linear polarization $|\phi_{top}\rangle$, with an angle following roughly $2\alpha$, emitting upwards without interacting with the rest of the metasurface, and (ii) the *indirectly* scattered state that emits upward after two passes through the metasurface. This latter component has the form $e^{2i\Phi}J^2|\phi_{top}\rangle$, where $J$ is the Jones matrix of a single-pass through the metasurface (controlled by the local pillars) and $\Phi$ is the propagation phase through the oxide slab. When $\Phi = \frac{\pi}{2}$, the scattering from the bottom interface is suppressed. By choosing $\theta$ and $\alpha$ such that the direct and indirect components from the top interface are orthogonal, CP emission is possible (Fig. 1(c)). More generally, any elliptical state is accessible by varying these parameters (Fig. 1(d)).

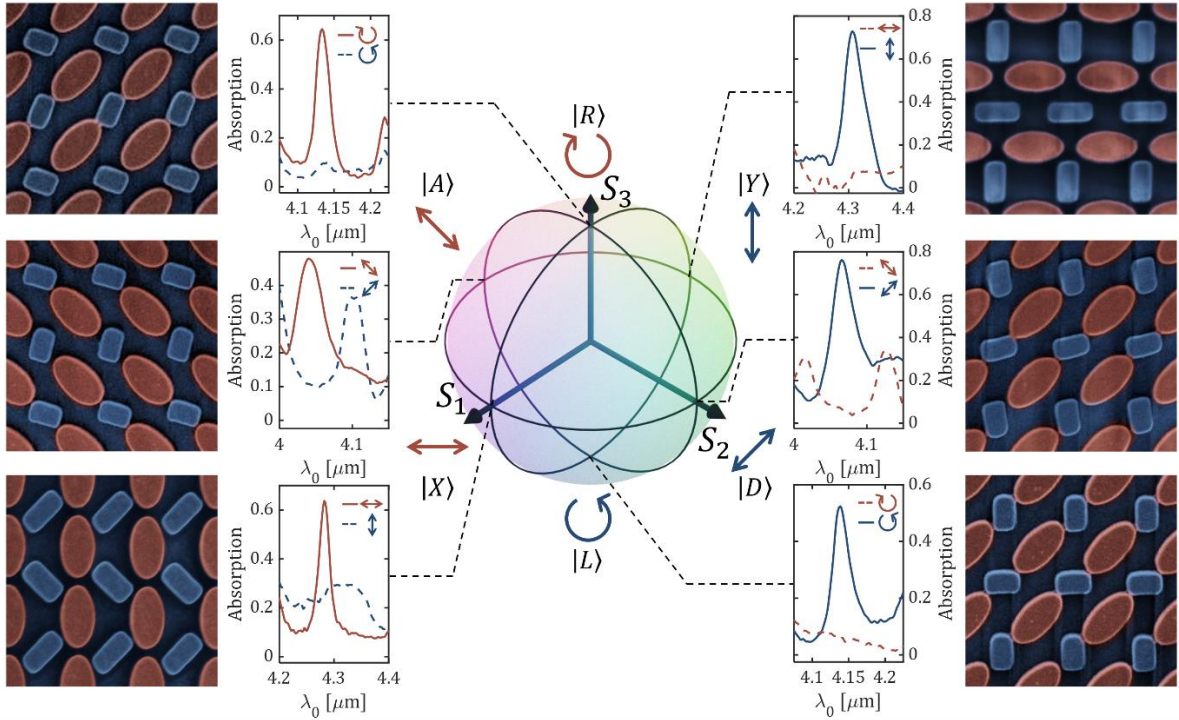

**Fig. 2. Experimental demonstration of polarization control.** Counterclockwise from the top left: polarization-resolved spectra from six metasurfaces designed for RCP-, antidiagonal-, x-, LCP-, diagonal-, and y-polarized emission. Insets: SEM images of the corresponding fabricated metasurfaces, with local elements shaded red and nonlocal elements shaded blue.

Figure 2 demonstrates six example metasurfaces designed to thermally absorb and emit extremal values on the Poincare sphere: $x$- and $y$-polarization ($|H\rangle$ and $|V\rangle$), diagonal and antidiagonal ($|D\rangle$ and $|A\rangle$), and RCP and LCP ($|R\rangle$ and $|L\rangle$). We show in the figure their polarization-dependent absorption features, which due to time-reversal validate their polarization-selective thermal emission, confirming the ability of single-layer dielectric metasurfaces to produce narrowband emission/absorption of custom polarization. Emphatically, in contrast to trivial control of polarization by orienting the entire metasurface, the polarization control here applies with respect to a *fixed* lattice. This capability enables construction of lattices in which the *local* polarization state varies across the device aperture. If $\alpha$ is appropriately adjusted as $\theta$ is varied across 180°, we may maintain a CP state. In this case, the local orientation carries an associated geometric phase, which may be locally patterned arbitrarily without affecting the extended q-BIC mode (Fig. 1(e)).

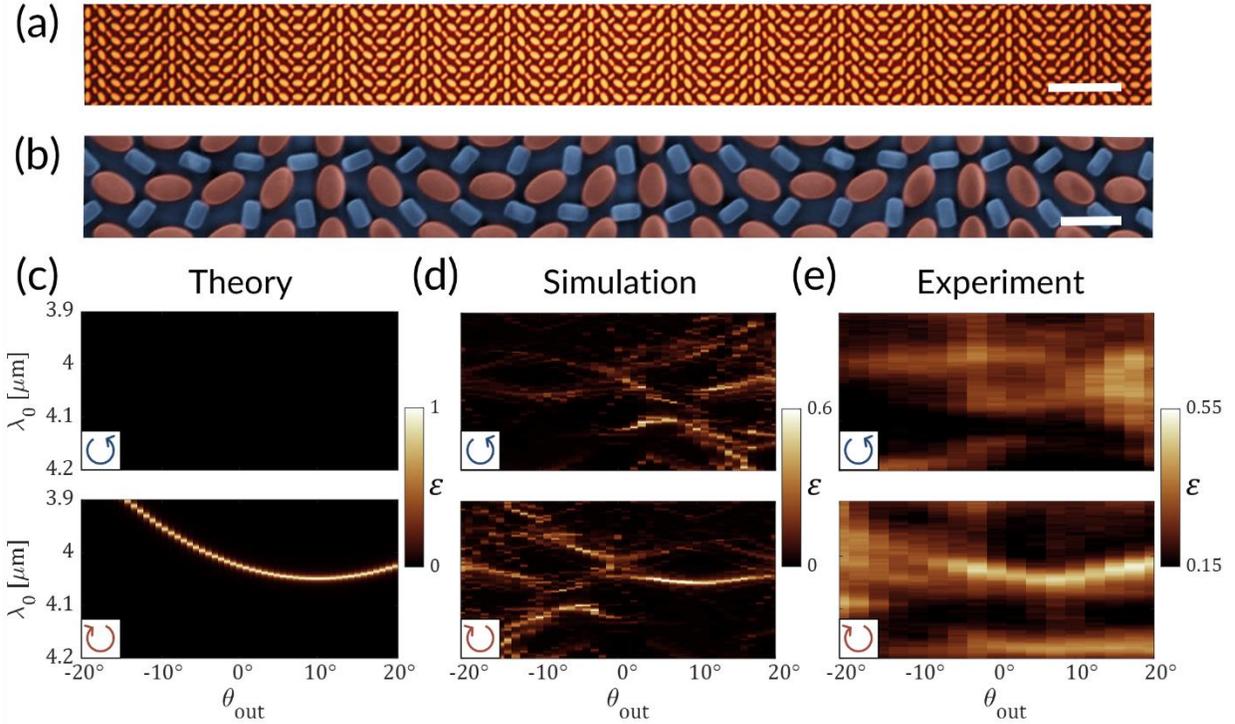

**Fig. 3. Spin-selective, unidirectional thermal emission.** (a) Optical and (b) scanning electron micrograph images of a fabricated thermal metasurface. Scale bars are $10\mu m$ and $2\mu m$, respectively. (c) Theoretical LCP (top) and RCP (bottom) emission from a unidirectional thermal metasurfaces, computed using STCMT. (d) Simulated LCP (top) and RCP (bottom) emission. (e) Measured LCP (top) and RCP (bottom) emission.

We leverage this degree of freedom to demonstrate a phase gradient profile of local phase applied to CP thermal emission. Figures 3(a) and 3(b) depict optical and scanning electron microscope images of the fabricated sample, with an aperture of $4 \times 3mm$ in lateral dimensions. Figure 3(c) shows the spin-dependent dispersion calculated through STCMT (detailed in the Supplementary Materials), an analytical model ideally suited to capture the dynamics of spatially varying q-BICs in nonlocal metasurfaces [28]. Here, the response is similar to what is shown in Fig. 1(c), but unidirectionally shifted in momentum. Our STCMT elegantly models the q-BIC mediated emission in spatially varying lattices in an idealized form that ignores nonidealities arising in practical implementations. Next, we use numerical simulations of a realistic metasurface designed based on the library in Fig. 1(e) to demonstrate this phenomenon in a concrete geometry (Fig. 3(d)), where additional spectral features associated with other lattice modes emerge. Notably, the targeted q-BIC and these additional modes are not completely orthogonal, resulting in modal coupling that manifests itself as spectral splitting in the dispersion. Intriguingly, this effect degrades away from the band edge, narrowing the range of wavelengths and angles across which the response is highly efficient. In turn, this phenomenon interestingly suggests a mechanism for enhancing the control over spatial coherence [29]. Finally, Fig. 3(e) showcases the experimental confirmation of the metasurface functionality, clearly showing a momentum-shifted narrow band with a single spin state, avoiding significant thermal emission of the opposite spin state to the opposite angle. These spin-selective, angle-resolved thermal emission measurements were performed at 300°C using a setup detailed in the Methods. The measured Q-factor ($Q \approx 100$) is lower than for the simulated structure, and the many unwanted additional spectral features are largely washed out. A small broadband background emissivity $\varepsilon \approx 15\%$ is present, but the targeted feature peaks at $\varepsilon \approx 55\%$. Future optimizations may further reduce the background and improve the peak prominence compared to this proof-of-principle demonstration.

In conclusion, we have experimentally introduced a thermal metasurface platform based on planarized, dielectric nanostructured thin films atop a gold mirror that absorb/emit designer polarization states within a narrow bandwidth in the infrared spectral region. In contrast to phonon-polariton or plasmonic approaches, our platform employs standard low-loss dielectrics structured via design principles rooted in symmetry; this approach is directly exportable to a wide range of optical frequencies and integrated platforms. The flexibility to specify arbitrarily the emitted state *locally* across a fixed lattice enables spin-selective unidirectional emission to a single half-space from an ultrathin film. With an appropriately designed phase profile, our platform may be extended to produce focused emission, orbital angular momentum and custom vector beams tailored across two spatial dimensions. As such, our results represent a major step towards a generalized solution to custom wavefront thermal generation in a compactified form factor, in which the metasurface does not need to be externally excited by a coherent source, but it leverages ubiquitous incoherent thermal oscillations. The wavelength-selective emission confers efficient use of thermal energy in comparison to blackbodies, which in conjunction with the prospect of custom wavefront generation, holds the promise for exciting opportunities in custom energy harvesting systems such as solar thermal electric generators. Despite the fundamentally incoherent nature of the underlying mechanism (incandescence), we have shown that suitably engineered structures leveraging both local and nonlocal optical scattering may corral the random fluctuations to radiate to a specific far field channel of choice. We stress that our results are fully compatible with reciprocity, and abide the universal modal radiation laws [30]. As a by-product, given that our thermally emitted beams carry a net transverse and spin momentum, we expect the metasurface to experience a backaction as it thermally emits. Overall, our proposed thermal metasurfaces fill a long-standing hole in our fundamental control of light, demonstrating asymmetric emission in both direction and polarization. Going forward, these concepts may advance the capabilities in related systems based on luminescence and fluorescence (e.g., light-emitting diodes and 2D materials), as well as coherent mechanisms such as lasing, nonlinear, and quantum media.

**Acknowledgements**

**Funding:** This work was supported by the Department of Defense Vannevar Bush Faculty Fellowship, and the Air Force Office of Scientific Research MURI program. This work was performed in part at the Advanced Science Research Center NanoFabrication Facility of the Graduate Center at the City University of New York. **Author contributions:** A.C.O. conceived the initial ideas. J.R.N and A.C.O designed the devices, performed numerical calculations, and analyzed the results. J.R.N and M.C. fabricated the samples. J.R.N. designed the experiments and characterized the samples. All authors wrote the manuscript. A.A. supervised the research. **Competing interests:** The authors declare no competing financial interest. **Data and materials availability:** All data needed to evaluate the conclusions in the paper are available in the main text or the supplementary materials.

## Materials and Methods

Materials

The thermal metasurfaces were fabricated using conventional semiconductor processing steps. To start, we used electron beam deposition to deposit 150 nm of Au and 40 nm of $Al_2O_3$ on top of an undoped, single-side-polished Si wafer. We then used plasma-enhanced chemical vapor deposition (PECVD) to deposit 2640 nm of $SiO_2$ and 1460 nm of Si. We also deposited individual layer of Si and $SiO_2$ in order to extract the dielectric functions using IR-variable angle spectroscopic ellipsometry (IR-VASE, Woollam). These dielectric functions were implemented into the numerical calculations. A schematic of the substrate is shown in Fig. S1(a) and dielectric functions of the Si and $SiO_2$ can be found below in Fig. S1(b) and S1(c), respectively.

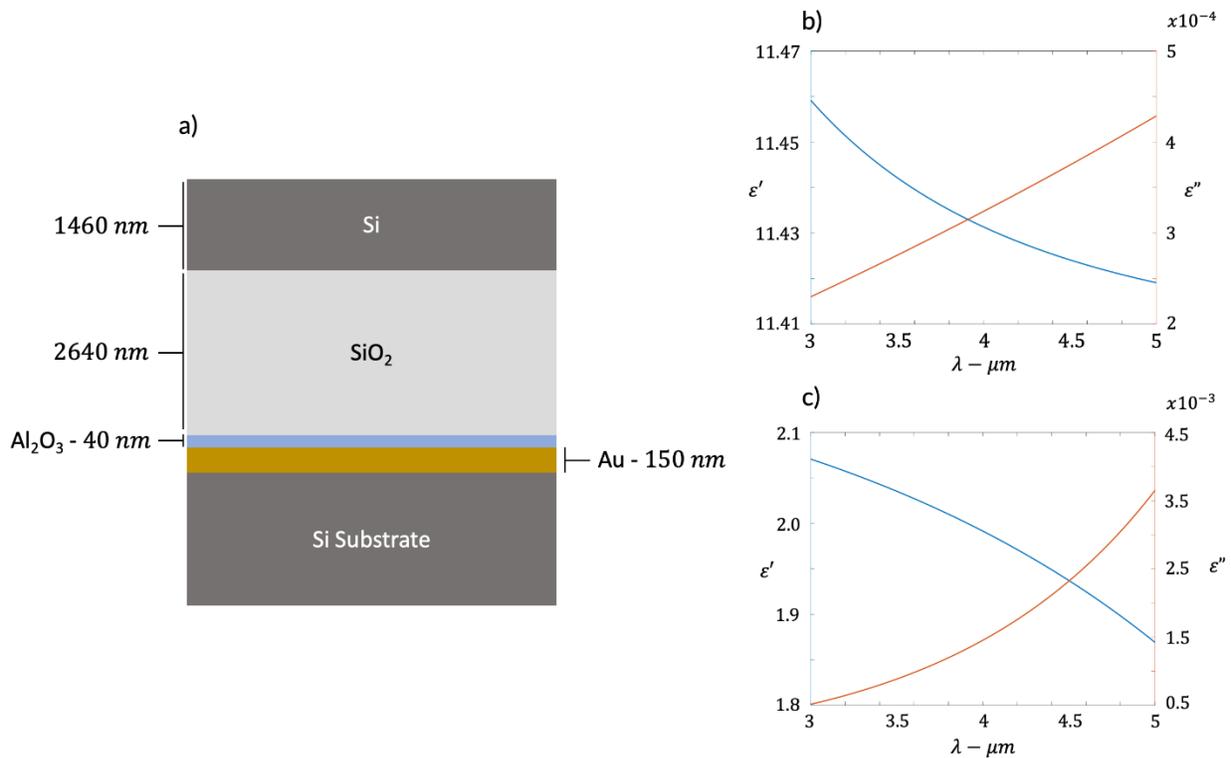

**Fig. S1.** (A) Schematic side view of pre-patterned substrate. Dielectric functions of (B) Si and (C) $SiO_2$.

The wafer was then diced into 1cm x 1 cm pieces, which were spin-coated with polymethyl methacrylate (PMMA). A 10 nm Au charge compensation layer was then deposited on top of the PMMA using sputtering and the samples were patterned using standard electron beam lithography (EBL). The samples were then developed using methyl isobutyl ketone (MIBK) and 40 nm of $Al_2O_3$ was deposited on top as an etch mask. A lift-off procedure was then performed using Remover PG and the substrates were etched using a Fluorine based inductively coupled plasma etcher (Oxford PlasmaPro System 100 Cobra).

Methods

Absorption and thermal emission measurements were performed using a Bruker INVENIO-R FTIR coupled with a Hyperion microscope or in-house built angle-resolved thermal emission setup.

Poincaré mapping of the absorption (Fig. 2 from main text) was performed using the Hyperion microscope. A 15x Cassegrain objective (Newport NA = 0.4) was used for both absorption and thermal emission measurements, which provides an azimuthal angle of incidence of $\theta \approx 22^o$ for all radial angles ($\phi = 0 - 180°$). To achieve a pure polarization state for the incident light, we first restricted the radial angles to only allow for light to be incident along the $x$-direction (see Fig. S2 below) using a 3D-printed aperture. This technique has been employed in previous publications as a way of restricting to only one polarization state and tuning the measured angle of incidence (31). Polarimetry was performed by placing a broadband mid-IR ($2.5 - 7\ \mu m$, Bernhard Halle Nachfl.) quarter waveplate (QWP) and $CaF_2$ wire grid polarizer (Thorlabs WP50H-C) in the beampath between the FTIR and the microscope.

To produce linear polarization states, the polarizer was placed in the polarizer slot of the microscope. The equator can then be mapped by changing the polarizer angle between 0° and 180°. Therefore, the S1 axis of the Poincare sphere can be isolated at polarizer angles of 0° and 90° and the S2 axis at polarizer angles of 45° and 135°. An alternate setup was required to produce CP polarization states, wherein the polarizer and QWP were placed in the beampath between the FTIR and the microscope as shown in Fig. S2.

Angle-resolved thermal emission measurements (Fig. 3 from main text) were performed using a Linkam THMS600 heated stage mounted upright on a commercially-available, motorized rotation stage (Thorlabs HDR50). The sample was heated to 300°C and the emission was collimated using a 90° off-axis parabolic mirror (OAP) and then sent into the FTIR. A broadband mid-IR QWP and $CaF_2$ wire grid polarizer were placed between the mirror and the FTIR to determine the polarization state of the thermal emission. The beampath of the thermal emission setup, showing the placement of the QWP and polarizer, is provided in Fig. S3. An array of vertically-aligned carbon nanotubes (500 $\mu m$ tall CNT on Si substrate, NanoLab) was used as a 'blackbody' reference sample in calculating the sample emissivity.

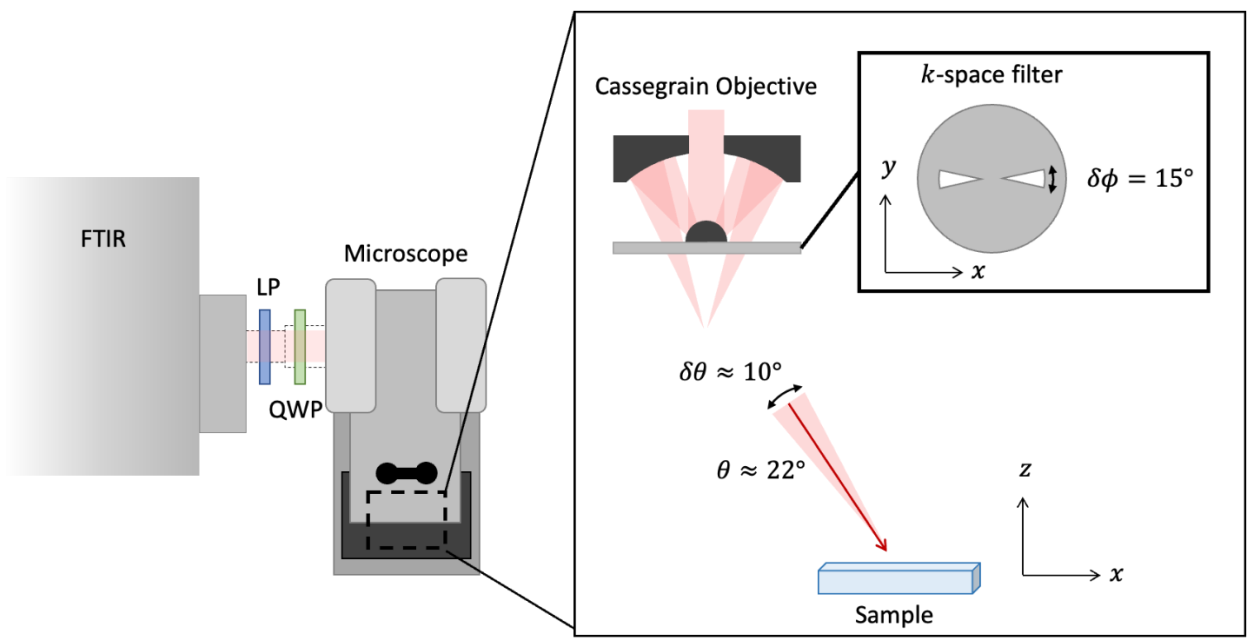

**Fig. S2. Setup for Fig. 2 of the main text.** (Left) Top-view schematic of Poincare mapping microscope setup showing the placement of the linear polarized (LP) and QWP used to generate incident RCP and LCP states. (Right) Side-view schematic of Cassegrain objective showing the placement of the $k$-space filter (inset).

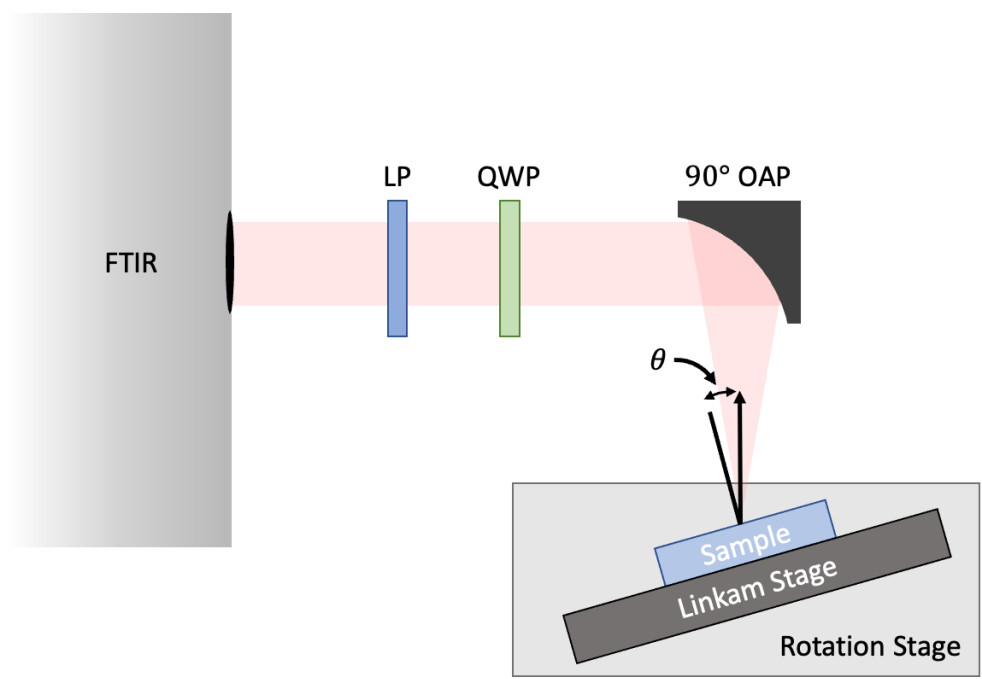

**Fig. S3.** Top-view schematic of angle-resolved thermal emission setup.

# Supplementary Text

## S.1 Coupled mode theory models

Q-BICs in periodic structures are well-captured by temporal coupled mode theory (TCMT). In Section S.1a, we developed a TCMT (32) on the basis of the idealized phenomenology of our design, providing a simple model for studying the spectral, polarization, and phase properties of the meta-units for thermal metasurfaces. However, in spatially varying metasurfaces, TCMT is not a suitable tool. Instead, a spatiotemporal coupled mode theory (STCMT), capturing both spectral and spatial information of the metasurface, is required (28). In Section S.1b, we develop a thermal STCMT on the basis of the meta-units from the developed TCMT, and reproduce the idealized functionalities of our thermal metasurfaces.

## S.1a Temporal coupled mode theory of the meta-units

TCMT begins with the dynamical equations

$$\frac{da(t)}{dt} + i(\omega_r + i\gamma_r + i\gamma_{nr})a(t) = \langle k | s_+(t) \rangle \tag{1}$$

$$|s_-(t)\rangle = C|s_+(t)\rangle + a(t)|d\rangle. \tag{2}$$

These equations describe the q-BIC in terms of a complex modal amplitude $a(t)$ as a function of time $t$ that is independent of space, a resonant frequency $\omega_r$, a radiative decay rate $\gamma_r$, and a nonradiative decay rate $\gamma_{nr}$. In Eqn. (1), the right-hand side describes the coupling into the q-BIC via incoming waves $|s_+(t)\rangle$ (normalized so that the norm squared is the incoming intensity) and the incoming coupling coefficients $|k\rangle$. In Eqn. (2), we compute the outgoing waves $|s_-(t)\rangle$ (normalized so that the norm squared is the outgoing intensity) in terms of the incoming waves, the direct scattering matrix $C$, and the outgoing coupling coefficients $|d\rangle$. Here, these kets represent column vectors having two entries, the first pertaining to incoming/outgoing waves with $x$ polarization and the second to waves with $y$ polarization. To satisfy conservation of energy, reciprocity, and time reversal invariance, the elements of the dynamical equations must satisfy **Error! Reference source not found.**,

$$|k\rangle = |d\rangle \tag{3}$$

$$\langle d|d\rangle = 2\gamma_r \tag{4}$$

$$C|d^*\rangle = -|d\rangle \tag{5}$$

TCMT proceeds by assuming by assuming time harmonic solutions [$\exp(-i\omega t)$], solving for the complex modal amplitude in Eqn. (1) and inserting the result into Eqn. (2) to yield the scattering equation:

$$|s_-(\omega)\rangle = S(\omega)|s_+(\omega)\rangle \tag{6}$$

with

$$S(\omega) = C + \frac{|d\rangle\langle d^*|}{i(\Omega - \omega)} \tag{7}$$

where $\Omega = \omega_r + i\gamma_r + i\gamma_{nr}$. Finally, a parameterization of the system under study, and consistent with Eqns. (3)-(5) is determined. Here, we are interested in a thermal metasurface having a lossless local response with full control over the unitary reflected Jones matrix:

$$C = e^{i\Phi_c} \text{Rot}[\theta] \begin{bmatrix} e^{i\Delta/2} & 0 \\ 0 & e^{-i\Delta/2} \end{bmatrix} \text{Rot}[-\theta] \tag{8}$$

where $\Phi_c$ is a reference phase, $\Delta$ is retardance between $x$ and $y$ polarizations, and the rotation matrix is

$$\text{Rot}[\theta] = \begin{bmatrix} \cos(\theta) & -\sin(\theta) \\ \sin(\theta) & \cos(\theta) \end{bmatrix} \tag{9}$$

We are interested in responses capable of full chirality, which requires the halfwave plate behavior $\Delta = \pi$. In this case, the background scattering matrix takes the simplified form

$$C = e^{i\Phi_c} \begin{bmatrix} \cos(2\theta) & \sin(2\theta) \\ \sin(2\theta) & -\cos(2\theta) \end{bmatrix}, \tag{10}$$

where we absorb all phase factors into $\Phi_c$ without loss of generality.

Last, a form of $|d\rangle$ consistent with Eqns. (5) and (10) while matching physical parameters of our system must be found. The principle behind the design of the thermal metasurfaces is depicted in Fig. S4(A), in which scattering induced by the perturbation originates from both the top and bottom interfaces of the silicon layer. The scattering from the top and bottom interfaces are linear polarizations, $|\phi_{top}\rangle$ and $|\phi_{bot}\rangle$, respectively

$$|\phi_{top,bot}\rangle = \begin{bmatrix} \cos(\phi_{t,b}) \\ \sin(\phi_{t,b}) \end{bmatrix}. \tag{11}$$

Then, each component scatters to free space via two routes, in which they either (directly) scatter upward immediately or (indirectly) first scatter downward [see Fig. S4(A)]. For $|\phi_{top}\rangle$, the sum of the direct and indirect processes may be written as

$$A_1|\phi_{top}\rangle + A_2 C|\phi_{top}\rangle \tag{12}$$

where $A_1$ and $A_2$ are weighting coefficients for each scattering event. Similarly, for $|\phi_{bot}\rangle$ the sum is

$$B_1 C_1|\phi_{bot}\rangle + B_2 e^{2ink_0 d} J|\phi_{top}\rangle. \tag{13}$$

where the phase factor $2nk_0 d$ accounts for the roundtrip phase accumulation through the oxide layer, and $J$ is the single-pass scattering matrix of the silicon metasurface layer, given by

$$J = e^{i\frac{\Phi_c}{2}} \frac{1-i}{2} \begin{bmatrix} i + \cos(2\theta) & \sin(2\theta) \\ \sin(2\theta) & i - \cos(2\theta) \end{bmatrix}. \tag{14}$$

Hence, the total scattered light can be expected to have the form

$$|d\rangle = (A_1 + A_2 C)\begin{bmatrix} \cos(\phi_t) \\ \sin(\phi_t) \end{bmatrix} + (B_1 + B_2 e^{2ink_0 d}) J \begin{bmatrix} \cos(\phi_b) \\ \sin(\phi_b) \end{bmatrix}. \tag{15}$$

To simplify this relation, we may choose the oxide thickness such that $e^{2ink_0 d} = -1$. In that case, if $B_1 \approx B_2$ we are left primarily with the scattering from the top interface. If the direct and indirect scattering processes have similar magnitude, i.e., $A_1 \approx A_2$, we have the form

$$|d\rangle \propto \begin{bmatrix} \cos(\phi) \\ \sin(\phi) \end{bmatrix} + e^{i\Phi_c/2}\begin{bmatrix} \cos(2\theta-\phi) \\ \sin(2\theta-\phi) \end{bmatrix}, \tag{16}$$

where we drop the subscript $t$ on the polarization angle $\phi$. Then, if the phase factor is engineered to be $e^{i\Phi_c} = -i$, we may expect the form

$$|d\rangle \propto \begin{bmatrix} \cos(\phi) \\ \sin(\phi) \end{bmatrix} - i\begin{bmatrix} \cos(2\theta-\phi) \\ \sin(2\theta-\phi) \end{bmatrix}. \tag{17}$$

Enforcing both Eqn. (4) and Eqn. (5), we arrive at

$$|d\rangle = \sqrt{i\gamma_r}\left(\begin{bmatrix} \cos(\phi) \\ \sin(\phi) \end{bmatrix} - i\begin{bmatrix} \cos(2\theta-\phi) \\ \sin(2\theta-\phi) \end{bmatrix}\right) \tag{18}$$

With this, we have all the elements of the scattering matrix in Eqn. (7), and may study the expected responses as a function of $\theta$ and scattered polarization $\phi$. And in particular, we may parameterize $\phi$ with the geometric angle $\alpha$; for the coordinate system in the main text, and based on the selection rules, we have

$$\phi \approx 2\alpha + 90° \tag{19}$$

Finally, near the band-edge frequency, we may approximate the q-BIC's dispersion parabolically as

$$\omega_r = \omega_0 + \frac{b}{2}k^2 + i(\gamma_r + \gamma_{nr}), \tag{20}$$

where $b$ is the band curvature coefficient. This form assumes that the coupling lifetime is independent of incident angle over the region of k-space of interest.

Figure S4(B) shows the predicted emission for RCP and LCP light as a function of $\theta_{out}$ and wavelength for the case with $\alpha = 22.5°$ and $\theta = 0$. Figure S4(C) shows Figure 1(c) in the main text for comparison. In units such that $\omega_0 = 2\pi/\lambda_0$, the TCMT in Fig. S4(B) uses the values $\omega_0 = 1.554\,\mu m^{-1}$, $\gamma_r = \gamma_{nr} = 2\pi \times 10^{-4}\,\mu m^{-1}$, and $b = 0.0064\,\mu m^2$.

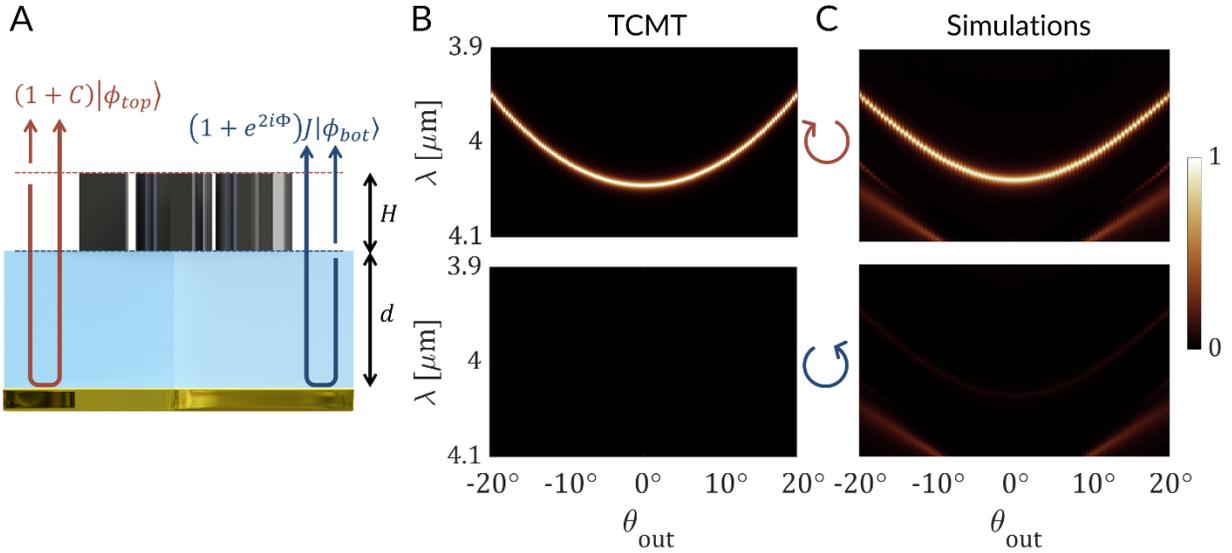

**Fig. S4. TCMT of periodic meta-units.** (A) Side-wave of geometry depicting the direct and indirect scattering components. (B) Emission predicted by TCMT for RCP (top) and LCP (bottom) for the fully chiral case with $\alpha = 22.5°$ and $\theta = 0$. (C) Emission of optimized RCP device from the main text.

Next, Fig. S5(A) reports the resonant polarization state as a function of $\alpha$ and $\theta$, predicted by the TCMT. Considering the 'toy model' nature of Fig. S4(A), the results show remarkably good qualitative agreement with the simulated results in Fig. S5(B) (reproduced from Fig. 1(d) from the main text). Notably, the discontinuity in longitude ($2\psi$) when precisely at a pole of the Poincare sphere ($2\chi = \pm\pi/2$) is observed only for the ideal TCMT case. In the simulated results, perfect chirality is not quite achieved, and these discontinuities are removed. Still, general agreement is observed in both the $2\psi$ and $2\chi$, confirming the underlying mechanism is well-captured by the toy model of Fig. S4(A). This also justifies extension of this TCMT to spatially varying lattices, carried out in the next section.

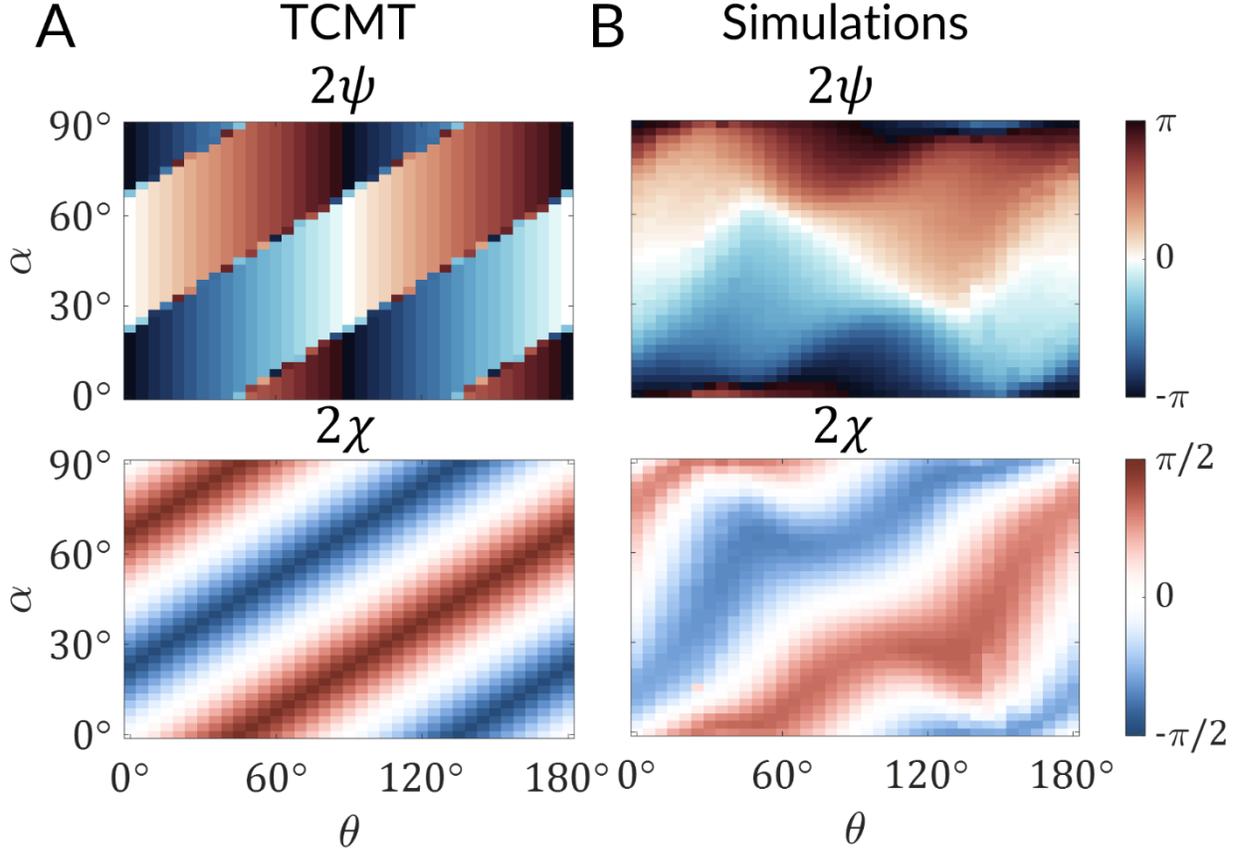

**Fig. S5. Polarization responses.** (A) Values of $2\psi$ and $2\chi$ of the resonant response as a function of $\alpha, \theta$ predicted by TCMT. (B) Simulated results from Fig. 1(d) in the main text for comparison.

### S.1b STCMT of a phase gradient metasurface

Now we seek an extension of the theory in the previous section that captures a metasurface in which the meta-units vary across the surface. This requires a STCMT that keeps the vectorial properties, adds loss, and is in reflection mode only.

We begin by assuming the position-dependent background scattering matrix $C(x)$ and coupling coefficients $|d(x)\rangle$ are *spatially instantaneous*, namely,

$$\begin{aligned} C(x) &\propto \delta(x-x') \\ |d(x)\rangle &\propto \delta(x-x') \end{aligned} \tag{21}$$

This amounts to assuming point-like perturbations that scatter the q-BIC perfectly locally. That is, adjusting from Eqn. (10), the background scattering simply follows

$$C(x) = \begin{bmatrix} \cos[2\theta(x)] & \sin[2\theta(x)] \\ \sin[2\theta(x)] & -\cos[2\theta(x)] \end{bmatrix}. \quad (22)$$

Meanwhile, the coupling coefficients follows (18) and reads:

$$|d(x)\rangle = \sqrt{\gamma_r} \left\{ \begin{bmatrix} \cos[\theta(x)] \\ \sin[\theta(x)] \end{bmatrix} - i \begin{bmatrix} \cos[2\theta(x) - \phi(x)] \\ \sin[2\theta(x) - \phi(x)] \end{bmatrix} \right\} \quad (23)$$

Under the above assumptions, it may be shown that the modal amplitude $a(x,t)$ follows a nonhomogeneous, non-Hermitian, effective mass Schrödinger equation:

$$\frac{da(x,t)}{dt} + i(\omega_0 + i\gamma_r + i\gamma_{nr})a(x,t) + ib\frac{d^2 a(x,t)}{dx^2} = \langle k(x)|s_+(x,t)\rangle, \quad (24)$$

while the outgoing waves follow, as usual:

$$|s_-(x,t)\rangle = C(x)|s_+(x,t)\rangle + a(x,t)|d(x)\rangle. \quad (25)$$

STCMT proceeds by assuming time harmonic solutions [$\exp(-i\omega t)$], solving Eqn. (24) via the Green's function method, and then inserting into Eqn. (25). This gives

$$|s_-(x,\omega)\rangle = \int dx' \sigma(x,x',\omega)|s_+(x,\omega)\rangle \quad (26)$$

with the scattering kernel

$$\sigma(x,x',\omega) = C(x)\delta(x-x') + G(x,x',\omega)|d(x)\rangle\langle d^*(x')|. \quad (27)$$

Here, the Green's function is

$$G(x,x',\omega) = -\frac{i}{b}\xi(\omega)\exp\left(-\frac{|x-x'|}{\xi(\omega)}\right) \quad (28)$$

parameterized by the complex nonlocality length (23)

$$\xi(\omega) = \sqrt{\frac{ib/2}{\gamma_r + \gamma_{nr} + i(\omega_0 - \omega)}}. \tag{29}$$

For clarity, we may separate the scattering kernel into its local and nonlocal components:

$$\sigma(x, x', \omega) = \rho_{loc}(x) + \rho_{nonlocal}(x, x', \omega) \tag{30}$$

where

$$\rho_{loc}(x) = C(x)\delta(x - x') \tag{31}$$

and

$$\rho_{nonlocal}(x, x', \omega) = G(x, x', \omega)|d(x)\rangle\langle d^*(x')|. \tag{32}$$

To aid in calculation, we note that we may compute the scattering in Eqn. (26) by casting these in discrete matrix form at each frequency:

$$|s_-\rangle = \sigma_\omega |s_+\rangle. \tag{33}$$

At a given frequency, $|s_+\rangle$ and $|s_-\rangle$ are column vectors with $2N$ elements corresponding to 2 polarizations and $N$ discrete position, and $\sigma_\omega$ is an $2N \times 2N$ matrix correlating the input positions $x'$ to output positions $x$ for each polarization combination. The absorption of a given input wave can be simply computed as

$$A = 1 - \langle s_+ | \sigma_\omega^* \sigma_\omega | s_+ \rangle. \tag{34}$$

The corresponding emission is to the time-reversed wave, most generally following the universal modal radiation laws (30)[30]. We must also consider the boundary conditions of the matrix $\sigma_\omega$; here we simply use radiative boundary conditions, implemented by default by truncating the matrix a certain position, though periodic boundary conditions may also be implemented (28).

Figure S6(A) shows, over a spatial span of $P$, the matrix forms of $\rho_{loc}$ (which is pure real), and the magnitude and phase of $\rho_{nonlocal}$ for a critically coupled ($\gamma_r = \gamma_{nr}$) phase gradient thermal metasurface emitting to an angle $k_g = 2\pi/P$, where $P = 12a$ and $a = 1.93\mu m$ is the lattice pitch. Figure S6(B) shows the same but for a device that spans a distance $W = 500\mu m$. The spatial resolutions are $dx = 0.5 \mu m$ and $dx = 1 \mu m$, respectively. Each matrix has four quadrants, corresponding to the polarizations of the input and output light (e.g., $C_{11}$ is the scattering of $x$ to $x$); within each quadrant is an $N \times N$ matrix. In the case of $\rho_{loc}$, the Dirac delta function means population solely of the main diagonal of each quadrant. Along the main diagonals are the values according to Eqn. (22), where here we use

$$\theta(x) = k_g x \tag{35}$$

Then, in the magnitude of $\rho_{nonlocal}$, we observe that the nonlocal correlation of optical energy across the surface is strongest when $x = x'$ (i.e., the main diagonals), but is nonzero away from this condition. The phase of the nonlocal scattering kernel encodes the phase gradient along its main diagonal. This kernel is computed using Eqn. (32), using

$$\rho_{nonlocal} = \begin{bmatrix} \rho_{11} & \rho_{12} \\ \rho_{21} & \rho_{22} \end{bmatrix} \tag{36}$$

$$\rho_{ij} = G_\omega D_{ij} \tag{37}$$

$$G_\omega = -\frac{dx}{b} \xi(\omega) \exp\left(-\frac{|X - X'|}{\xi(\omega)}\right) \tag{38}$$

$$D_{ij} = |d_i\rangle\langle d_j^*| \tag{39}$$

$$|d_1\rangle = \sqrt{\frac{\gamma_r}{2}} \left[\cos(\phi) - i\cos(2\theta - \phi)\right] \tag{40}$$

$$|d_2\rangle = \sqrt{\frac{\gamma_r}{2}} \left[\sin(\phi) - i\sin(2\theta - \phi)\right] \tag{41}$$

where we use the position matrices

$$\begin{aligned} \mathbf{x} &= \begin{bmatrix} x_1 & x_2 & \dots & x_N \end{bmatrix}^T \\ X &= \begin{bmatrix} \mathbf{x} & \mathbf{x} & \dots & \mathbf{x} \end{bmatrix} \\ X' &= X^T \end{aligned} \tag{42}$$

In particular, in order to maintain the circular polarization we compute Eqns. (36)-(41) with

$$\phi(x) = \theta(x) + \pi/4. \tag{43}$$

In this way, we compute the emission of both circular polarizations, shown in Fig. S6(C) and (D). We see that the band-edge has been shifted asymmetrically solely for one of the spins, while the opposite spin has no features.

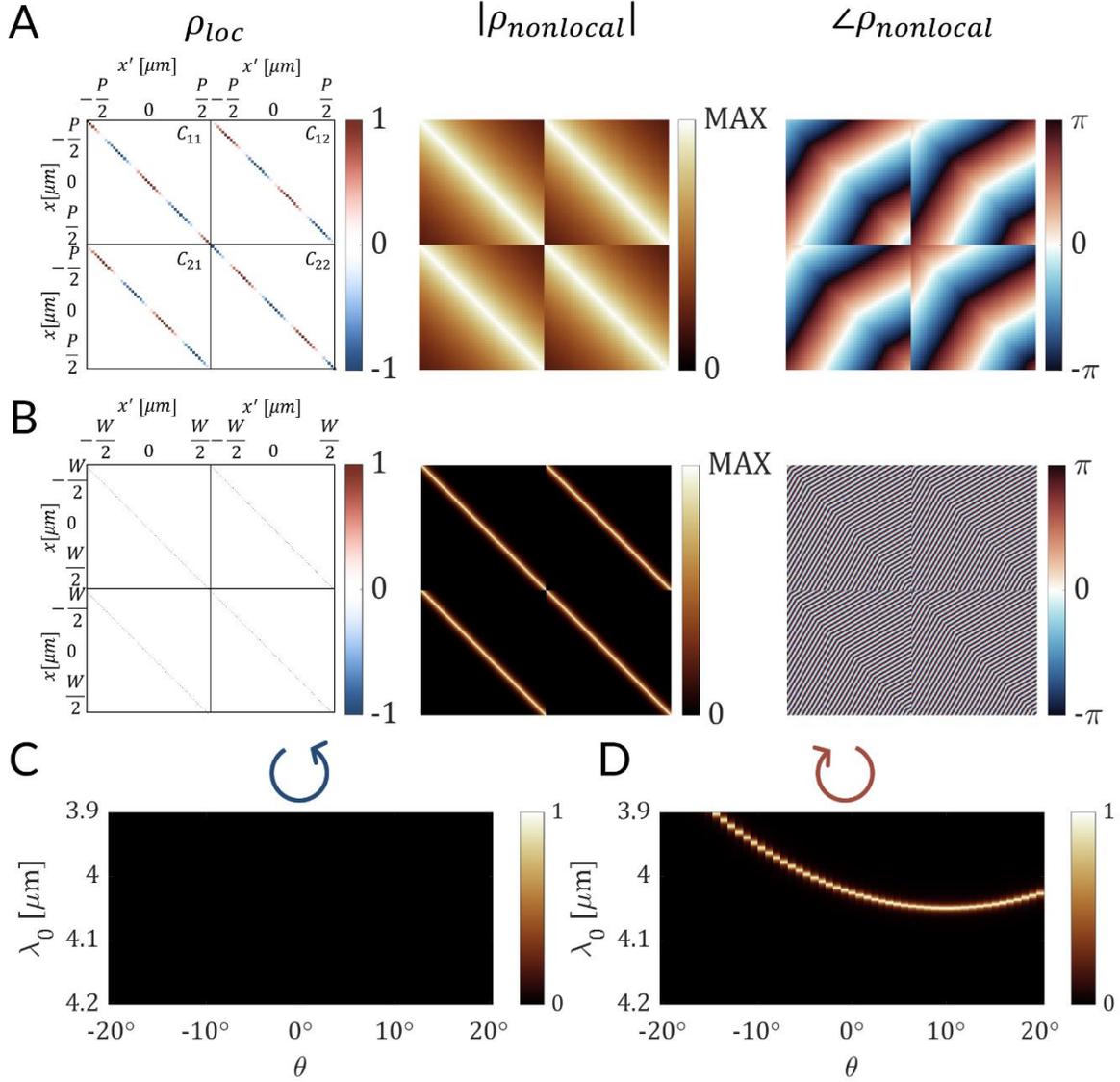

**Fig. S6. STCMT prediction of asymmetric emission**. (A) Elements of the scattering kernel for a phase gradient $k_g = 2\pi/P$ for a device with a single period $P = 12a$, where $a = 1.93\mu m$. (B) Elements for the same phase gradient but over a device spanning $W = 500\mu m$. (C,D) LCP and RCP emission computed using the elements in (B).

## S.2 Angle-tunability of band-edge emission

Due to the *p*2 space group of the dimer, $2\pi$ coverage of the geometric phase ($\Phi_{geo}$) is achieved by rotating both the local and nonlocal elements. The resultant spin-dependent shift in the emission angle can be calculated from the generalized Snell's Law (15):

$$k_0 \sin(\theta_\sigma) = \sigma m \frac{\partial \Phi_{geo}}{\partial x} \tag{44}$$

Where $m$ is the diffraction order, $\sigma = \pm 1$ expresses the handedness-dependence of the incident/emitted light, $\theta_\sigma$ is the angle of band-edge emission, and $k_0$ is the wavenumber in free-space. We know that the phase gradient achieves $2\pi$ phase coverage over one 'supercell' of the phase gradient, which allows for efficient coupling to the $m = 1$ diffractive order. Therefore, we can solve for the emission angle in terms of the unit cell geometry.

$$\theta_\sigma = \sin^{-1}\left(\frac{\sigma}{k_0} \frac{\partial \phi_{geo}}{\partial x}\right) = \sin^{-1}\left(\frac{2\pi\sigma}{k_0 a N}\right) \tag{45}$$

Where $a$ and $N$ are the period of a single unit cell and the number of cells in one supercell, respectively. In Fig. 3 of the main text we show unidirectional RCP emission from of a device with a supercell length of $N = 12$. The spin-corrected shift in $k_\parallel$ results in the band edge emission of RCP shifting to $\theta_{RCP} = +10°$. Note that in the absence of birefringence, this would reciprocally result in the shifting of LCP emission to $\theta_{LCP} = -10°$, however the device here is designed to couple exclusively to RCP radiation. To demonstrate that the angle of band edge emission can be tuned by changing the length of the supercell, two additional samples were fabricated with $N = 8$ and $N = 16$. The measured angle of band edge emission is shown in Fig. S7 showing good agreement with simulations (CST Studio Suite) and Eqn. (45).

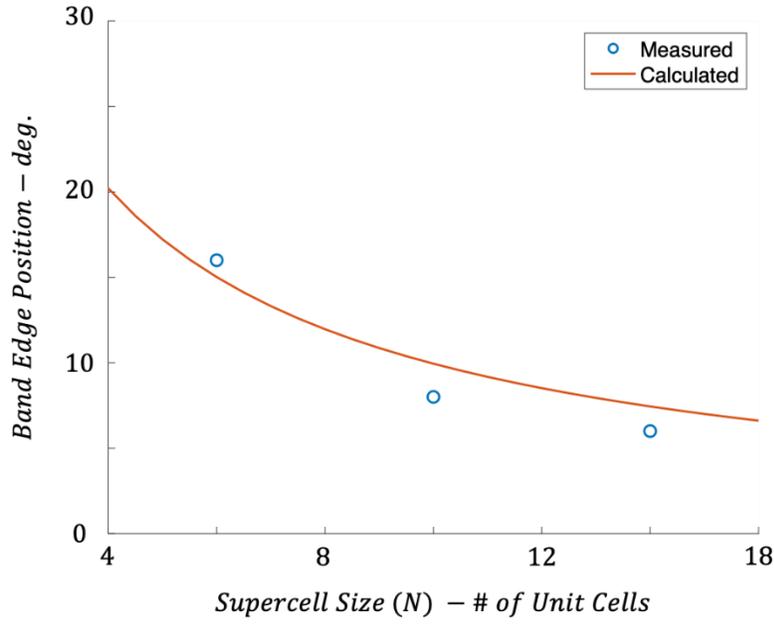

**Fig. S7.** Band edge emission angle tunability by altering supercell size.

## S.3 Dispersion along the dimerization direction

The dispersion of the thermally populated q-BIC supported by a periodic thermal metasurface can be approximated by a Taylor expansion about $\theta = 0$, as seen in Eqn. (20). Here, the coefficient $b$ is a phenomenological measure of the band curvature. Knowing the degree of band curvature, along with the lifetime of the mode ($\tau$), the spatial coherence length can be approximated as $L_c \approx \sqrt{b\tau}$ (23). Therefore, the larger the band curvature the greater the spatial coherence at the band edge.

In Fig. S8, we provide the dispersion along the x-direction (orthogonal to the dimerization direction) and y-direction (along the dimerization direction) of a periodic thermal metasurface. Notably, the band curvature along the y-direction is shown to be much larger than along the x-direction. Given that the band edge emission is at $\theta_x = \theta_y = 0$ and therefore the linewidth is identical in both cases, this implies that the spatial coherence along the y-directions is larger that along the x-direction.

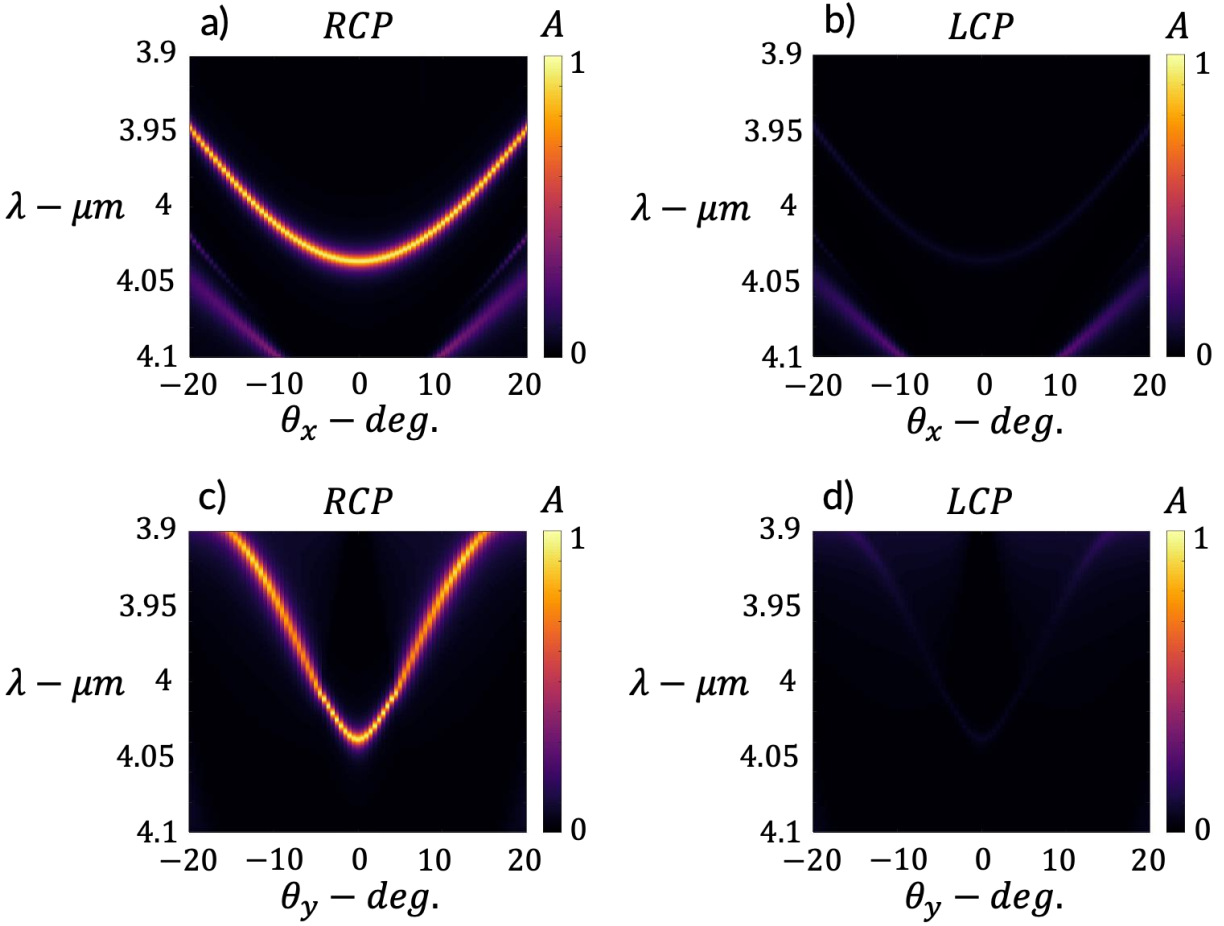

**Fig. S8:** Band edge emission angle tunability by altering supercell size.

## S.4 Wavelength tuning of the response

The resonant wavelength of the QBIC-supporting thermal metasurface is dependent on all of the geometric parameters of the unit cell: $D_t$, $H$, $t_{SiO_2}$, $p$, $L$, and $W$. In this work, we have set these parameters such that absorption/emission at 4 $\mu m$ is achieved. Therefore, in the spectral range of low dispersion for Si and SiO$_2$ the resonant wavelength can be tuned and polarization selectivity can be maintained by simply scaling these geometric parameters. This is demonstrated below where we show that the absorption peak can be blue- or redshifted by simply scaling the geometric parameters of the structure. Further, as the coupling to QBIC modes arise from symmetry breaking in dielectric photonic crystal slabs, the analysis provided here translates readily to any other system composed of isotropic materials with low dispersion.

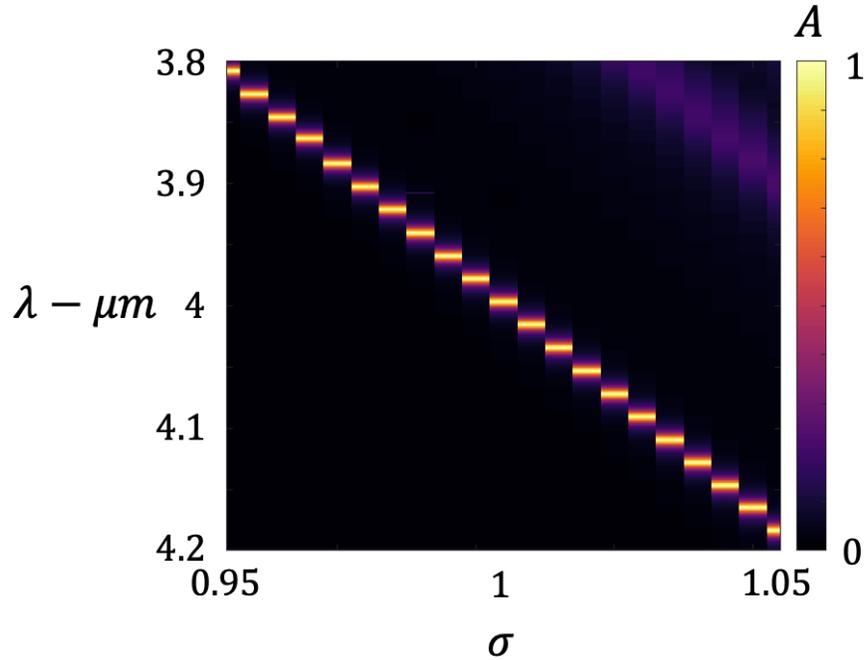

**Fig. S9.** Calculated absorption of thermal metasurface designed to selectively absorb $y$-polarized light, showing the shift in resonant wavelength ($\lambda$) as function of the scaling factor ($\sigma$). The scaling factor acts as a linear multiplier on the geometric parameters ($H$, $d$, $D_t$, $p$, $L$, and $W$).

### S.5 Temperature dependence of the emission

We varied the temperature and monitored the band edge emission $(\theta, \phi) = (10°, 0°)$ of the unidirectional emitter in Fig. S3 to determine the temperature stability of the thermally populated q-BIC. As seen in Fig. S10, the increased temperature results in a slight increase in the emissivity and redshift of the resonant wavelength, demonstrating that the emission is robust against temperature fluctuations. This can be attributed to the temperature dependence of the imaginary permittivity of Si and SiO$_2$ (33), increasing the losses and altering the real permittivity through the Kramer's-Kronig relation.

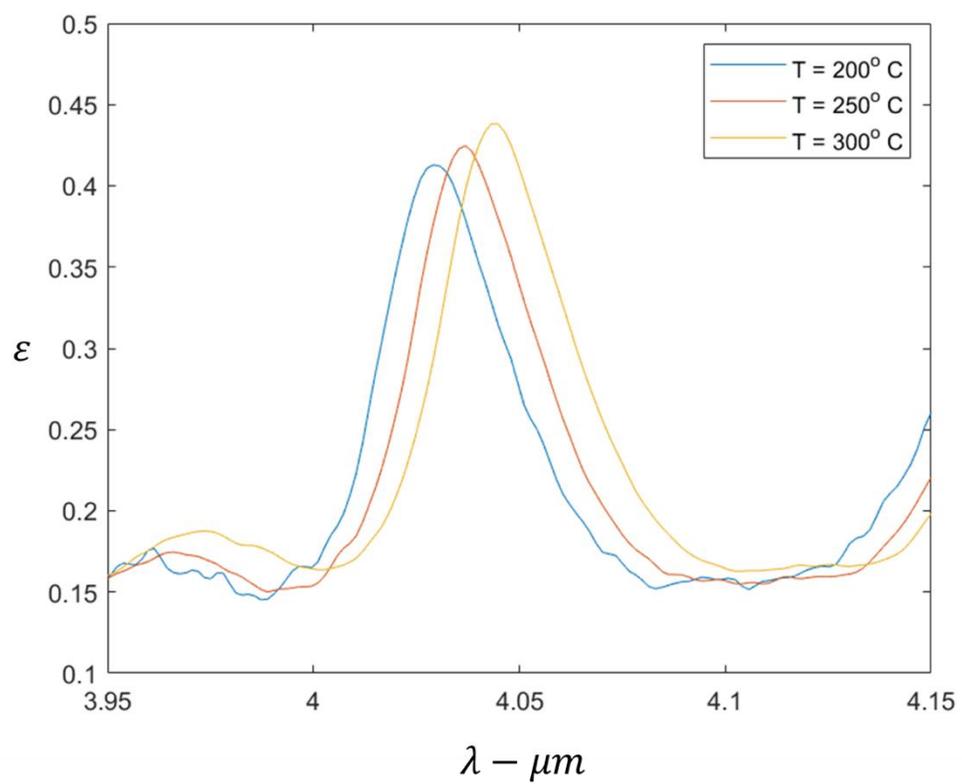

**Fig. S10.** Temperature dependence of the unidirectional emitter band edge emission.